# Discriminative Robust Deep Dictionary Learning for Hyperspectral Image Classification

Vanika Singhal, Hemant K. Aggarwal, Snigdha Tariyal and Angshul Majumdar, *Senior Member, IEEE*

*Abstract*—This work proposes a new framework for deep learning that has been particularly tailored for hyperspectral image classification. We learn multiple levels of dictionaries in a robust fashion. The last layer is discriminative that learns a linear classifier. The training proceeds greedily; at a time a single level of dictionary is learnt and the coefficients used to train the next level. The coefficients from the final level are used for classification. Robustness is incorporated by minimizing the absolute deviations instead of the more popular Euclidean norm. The inbuilt robustness helps combat mixed noise (Gaussian and sparse) present in hyperspectral images. Results show that our proposed techniques outperforms all other deep learning methods – Deep Belief Network (DBN), Stacked Autoencoder (SAE) and Convolutional Neural Network (CNN). The experiments have been carried out on benchmark hyperspectral imaging datasets.

## I. INTRODUCTION

IN recent years deep learning has successfully solved decade old problems in speech and image processing. To understand the impact of deep learning it suffices to say that top level conferences (InterSpeech and CVPR / ICCV) in these areas have more than half of the papers on topics related to deep learning. The popularity of deep learning in image analysis have motivated researchers in hyperspectral imaging to explore these techniques.

The concepts of deep learning are not new, they have been known since the early days of neural networks from the 90's. Basically deep learning meant that instead of having a single hidden layer in a neural network, one can have multiple hidden layers. However there were two fundamental bottlenecks in early 90s that prevented the success of deep learning.

First, more layers in a neural network means more parameters (network weights) to learn; this in turn would require more data. In early 90's we did not have the provision to store such large volume of data required to train such deep networks. Limitations in memory was the first hindrance.

Second, we did not have enough computational power to train such large networks. It is only in the mid 2000's that deep learning was successful in penetrating a decade long barrier in speech recognition [1]. It was made possible only with GPUs. The computational power that can be garnered from parallel processing was lacking in the 90's.

Today it is generally accepted that there are three pillars of deep learning – stacked autoencoder (SAE), deep belief network (DBN) and convolutional neural network (CNN). All three have found straightforward application in hyperspectral imaging – [2] uses SAE, [3] uses DBN and [4] uses CNN. Even without handcrafting the input features, deep learning outperforms state-of-the art results obtained from domain expertise – the success of [2-4] corroborates the thought provoking discussion at Technion [5].

Deep learning extends beyond the realms of neural networks. It is a powerful representation learning tool. Before deep learning gained popularity in speech and signal analysis, researchers used two classes of features – 1. Hand crafted features based on classical computer vision techniques such as interest points [6, 7] or textures [8, 9]; 2. Statistical features based classical factor analysis [10-13]. Both required expertise and understanding of hyperspectral data. Deep learning techniques on the other hand do not require such expertise. Instead of 'designing' the feature extraction model it 'learns' the model given sufficiently large volume of data.

To distinguish between classical feature extraction / dimensionality reduction with such model learning, the term 'representation learning' is used instead. Basically these are the features obtained from the penultimate layer of a neural network. The learned representation need not be used with a neural network type classifier – its application is broader; one is free to choose any classifier on the learnt representation.

In between the stages of hand-crafted feature extraction and deep learning, dictionary learning gained popularity in image analysis and computer vision. There are some studies in hyperspectral image analysis on this topic as well [14-16]. These studies combined representation learning with classical discriminative factor analysis techniques.

In a recent work, we proposed a new tool for deep learning dubbed – deep dictionary learning (DDL) [17]. It is the first work showing how deep architectures can be built from greedy dictionary learning. In the just concluded WHISPERS workshop [18] we have shown how DDL it yields significantly superior results compared to SAE [2] and DBN

V. Singhal is with Indraprastha Institute of Information Technology, Delhi, India (e-mail: vanikas@iiitd.ac.in).
H. K. Aggarwal is with Indraprastha Institute of Information Technology, Delhi, India (e-mail: hemanta@iiitd.ac.in).
S. Tairyal is with Indraprastha Institute of Information Technology, Delhi, India (e-mail: snigdha1491@iiitd.ac.in).
A. Majumdar is with Indraprastha Institute of Information Technology, Delhi, India (e-mail: angshul@iiitd.ac.in).

[3] for hyperspectral image classification problems [18].

In deep dictionary learning, the basic idea is to learn multiple levels of dictionary in a greedy fashion, i.e. the first level learns a standard dictionary and coefficients. In subsequent levels the coefficients from the previous level acts as inputs for dictionary learning. Although it yields results better than well known deep learning tools, there is scope for improvement. Standard dictionary learning is based on minimizing the Euclidean norm; this is optimal when the noise is Gaussian. It is well known that hyperspectral images are corrupted by a mixture of Gaussian and sparse noise [19, 20]. The sparse noise arises from diffraction grating and transient dead pixels [21].

Ideally minimizing the Euclidean norm is optimal for Gaussian noise; for sparse noise one needs to minimize the $l_0$-norm. A compromise between these two extremes is minimizing the absolute deviations, the $l_1$-norm; a classical metric in robust statistics [22-24]. The prior study [17, 18] used the standard dictionary learning with $l_2$-norm cost function; in this work we propose the framework for robust deep dictionary learning (RDDL).

The second improvement is in the addition of discrimination penalty. The last level of dictionary maps onto the target labels. By learning the map, we can classify within the deep dictionary learning framework; we do not need a separate classifier like [17, 18]. This is in lines with other deep learning tools like stacked autoencoder and deep belief network where a soft-max classifier or logistic regression is learnt to complete the deep neural network.

The rest of the paper is organized into several sections. The following section discusses prior work. The proposed formulation is given in section 3. The experimental results are shown in section 4. The conclusions of this work and future direction of research is discussed in section 4.

## II. LITERATURE REVIEW

### A. Representation Learning

In recent times many a papers are being published on the topic of 'deep learning' in the context of hyperspectral imaging. Hence an elaborate discussion is redundant. We briefly discuss the different deep learning techniques for the sake of completeness.

Autoencoders (AE) and restricted Boltzmann machine (RBM) have been used to build deep learning architectures – stacked autoencoders and deep belief network respectively.

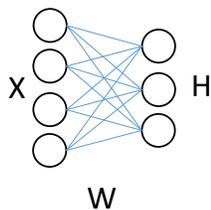

Fig. 1. Restricted Boltzmann Machine

Restricted Boltzmann Machine (RBM) [25] is a popular representation learning method; the schematic representation is shown in Fig. 1. RBM is an undirected graphical model. The objective is to learn the network weights ($W$) and the representation ($H$). This is achieved by optimizing the Boltzman cost function given by:

$$p(W,H) = e^{H^T WX} \qquad (1)$$

Here $X$ is the input data; the samples are stacked as columns.

Basically RBM learns the network weights and the representation / feature by maximizing the similarity between the projection of the input (on the network) and the features in a probabilistic sense. Since the usual constraints of probability apply, degenerate solutions are prevented. The traditional RBM is restrictive – it can handle only binary data. The Gaussian-Bernoulli RBM [26] partially overcomes this limitation and can handle real values between 0 and 1. However, it cannot handle arbitrary valued inputs (real or complex).

Deep Boltzmann Machines (DBM) [27] is an extension of RBM by stacking multiple hidden layers on top of each other (Fig. 2). The RBM and DBM are undirected graphical models. For training deep architectures, targets are attached to the final layer and fine-tuned with back propagation. Usually this is a soft-max classifier or logistic regression. Training of the classifier with the deep architecture completes the deep neural network.

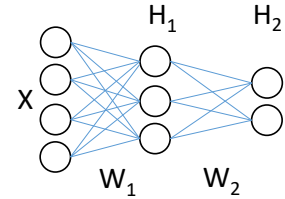

Fig. 2. Deep Botlzmann Machine

Another basic building block for training deep neural networks is autoencoder [28]. The architecture is shown in Fig. 3.

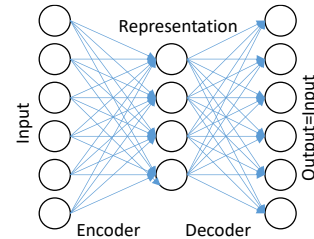

Fig. 3. Autoencoder

$$\min_{W,W'} \| X - W'\phi(WX) \|_F^2 \qquad (2)$$

The cost function for the autoencoder is expressed above. $W$ is the encoder, and $W'$ is the decoder. $\varphi$ denotes the non-linear activation function. The autoencoder learns the encoder and decoder weights such that the reconstruction error is minimized. Essentially it learns the weights so that the representation $\phi(WX)$ retains almost all the information (in the Euclidean sense) of the data, so that it can be reconstructed back. Once the autoencoder is learnt, the decoder portion of the autoencoder is removed and the target is attached after the representation layer.

To learn multiple layers of representation, the autoencoders are nested into one another. This architecture is called stacked autoencoder.

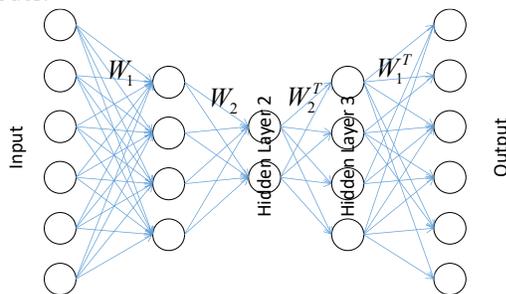

Fig. 4. Two-layer Stacked Autoencoder

For such a stacked autoencoder, direct optimization problem is complicated. The workaround is to learn the layers in a greedy fashion [29]. First the outer layers are learnt; and using the features from the outer layer as input for the inner layer, the encoding and decoding weights for the inner layer are learnt.

For training deep neural networks, the decoder portion is removed and targets attached to the inner layer. A soft-max or logistic regression cost function is used; the complete structure is then fine-tuned with backpropagation.

Autoencoders and RBMs are used for training generic neural networks. For problems arising in image processing, convolutional neural networks (CNN) are more popular. We will briefly explain CNN via the popular 2-layer LeNet architecture (Fig. 5).

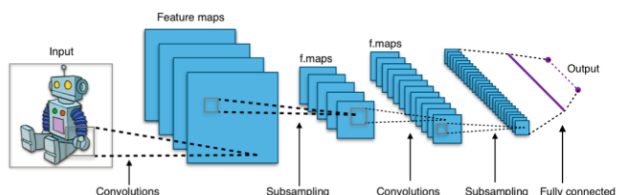

Fig. 5. LeNet Architecture

Owing to the local correlations in images, CNN has been widely successful in image analysis. The basic building block of a deep CNN is a convolution layer followed by a pooling layer. In the convolution layer, multiple convolutional kernels are applied on the image to generate the corresponding feature maps. Usually, only the positive portion of the output is retained from the rectified linear unit activation function (other activation functions like sigmoid and tanh are also used). This completes the convolution stage. Following the convolution is the pooling layer. In this stage, the obtained feature maps are sub-sampled by either taking the average or the maximum (other pooling techniques are less popular) from a pre-defined window. This leads to the pooled feature maps (f.maps in Fig. 5). This concludes one stage of the CNN.

For deeper architectures, for example the 2-layer LeNet, the same process if repeated on the pooled feature maps. After the second layer, the output is flattened and a fully connected layer is used to map it to the output targets. The entire architecture is learnt greedily via backpropagation.

B. *Deep Learning in Hyperspectral Imaging*

As mentioned in the introduction, stacked autoencoders [2] and deep belief networks have been used for image classification [3]. Both of them use a logistic regression for classification.

CNNs are also becoming popular in this area. However most CNN based results only vary slightly from one another in their configurations. For example in [4], each layer of CNN is independently trained. Instead of using the outputs as a fully connected layer (as in LeNet), they use support vector machine for classification.

In general CNNs are data hungry. When training samples are limited, they tend to overfit and yield poor results on testing. This is the reason the standard CNN architectures, popular in computer vision cannot be directly applied in hyperspectral imaging. In [30], multiple CNNs are trained randomly and their outputs are boosted. This is done to prevent overfitting and improve generalization.

A more recent work [31] employs the pre-training fine-tuning paradigm for CNN based hyperspectral image classification. They pre-train on large volume of semi-accurate data; and fine-tune the pre-trained model on accurately labeled data for final classification.

There are also several studies on fusing traditional features with deeply learnt representation. For example in [32], a deep autoencoder is used for feature extraction and a collaborative representation based classifier is used to enforce spatial correlations. In [33], traditional spectral features are extracted; separately CNN based spatial features are extracted. For classification, both of them are fused to form the final decision. An interesting work [34] introduces active learning into the deep learning framework; this has mainly to do with the limited availability of training samples. In [35], a simple voting strategy is used. A standard CNN is learnt; during classification, instead of estimating the class based on one pixel a voting is done based on the classes of adjacent pixels as well. Local redundancy in the image is thus captured.

III. DISCRIMINATIVE ROBUST DEEP DICTIONARY LEARNING

A. *Deep Dictionary Learning*

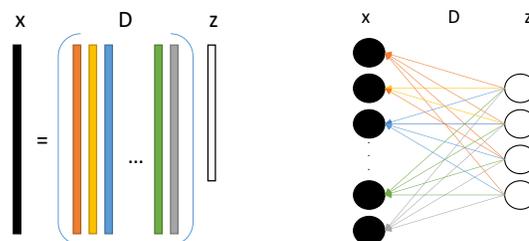

Fig. 6. Dictionary Learning. Left – Traditional Interpretation. Right – Our Neural Network Interpretation

The popular interpretation for dictionary learning is that it learns a basis ($D$) for representing ($Z$) the data ($X$) (see Fig. 6-

Left); for sparse coding, the representation needs to be sparse. The columns of *D* are called 'atoms'.

In this work, we have an alternate interpretation of dictionary learning. Instead of interpreting the columns as atoms, we can think of them as connections between the input and the representation layer (Fig. 6-Right). To showcase the similarity, we have kept the color scheme intact. Unlike a neural network which is directed from the input to the representation, the dictionary learning kind of network points in the other direction – from representation to the input. This is what is called 'synthesis dictionary learning' in signal processing.

The dictionary is learnt so that the features (along with the dictionary) can synthesize / generate the data. The formulation is:

$$X = DZ \qquad (3)$$

The dictionary and the coefficients are learnt by minimizing the Euclidean cost function:

$$\min_{D,Z} \|X - DZ\|_F^2 \qquad (4)$$

This formulation was introduced by Lee and Seung [36]. Today, most studies impose an additional sparsity constraint on the representation (*Z*) [37] but it is not mandatory.

Building on the neural network interpretation, [17] proposes deeper architecture with dictionary learning. An example of two-layer architecture is shown in Fig. 7.

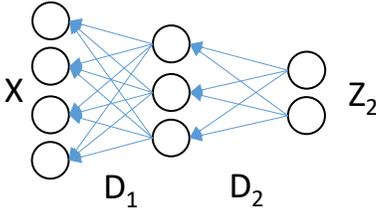

Figure 7. Deep Dictionary Learning

For the first layer, a dictionary is learnt to represent the data. In the second layer, the representation from the first layer acts as input and it learns a second dictionary to represent the features from the first level. This concept can be further extended to deeper layers. Deep dictionary learning can be used for representation learning. It requires a separate classifier. It has been used for hyperspectral image classification in [18].

Mathematically, the representation at the second layer (Fig. 7) can be written as:

$$X = D_1 \varphi(D_2 Z_2) \qquad (5)$$

Here $\varphi$ is the activation function. The activation function is absent in the first layer since *X* can take any real value; hence we do not want to use a function that squashes the output between 0 – 1 or -1 – +1. The activation functions prevents the dictionaries to collapse into a single level.

The challenges of learning multiple levels of dictionaries in one go are the following:
1) Recent studies have proven convergence guarantees for single level dictionary learning [38-40]. These proofs would be very hard to replicate for multiple layers.
2) Moreover, the number of parameters required to be solved increases when multiple layers of dictionaries are learnt simultaneously. With limited training data, this could lead to over-fitting.

Therefore DDL proposes to learn the dictionaries in a greedy manner which is in sync with other deep learning techniques. Moreover, layer-wise learning will guarantee the convergence at each layer. The diagram illustrating layer-wise learning is shown in Fig. 8.

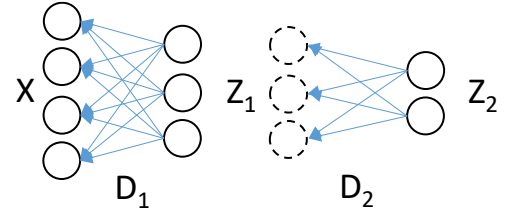

Fig. 8. Greedy Layer-wise Learning

Extending this idea, a multi-level dictionary learning problem with non-linear activation ($\varphi$) can be expressed as,

$$X = D_1 \varphi\big(D_2 \varphi(...\varphi(D_N Z))\big) \qquad (6)$$

Ideally, we would have to solve the following problem.

$$\min_{D_1,...D_N,Z} \|X - D_1 \varphi(D_2 \varphi(...\varphi(D_N Z)))\|_F^2 + \mu \|Z\|_1 \qquad (7)$$

However, such a problem is highly non-convex and requires solving huge number of parameters. With limited amount of data, it will lead to over-fitting. To address these issues, as mentioned before, DDL proposes a greedy approach where we learn one layer at a time. With the substitution $Z_1 = \varphi\big(D_2\varphi(...\varphi(D_N Z))\big)$, Equation (6) can be written as $X = D_1 Z_1$ such that it can be solved as single layer dictionary learning. The representation $Z_1$ is not sparse. Hence it can be solved using alternating minimization –

$$\min_{D_1, Z_1} \|X - D_1 Z_1\|_F^2 \qquad (8)$$

Optimality of solving (8) by alternating minimization has been proven. Therefore we follow the same approach. The dictionary *D* and the basis *Z* is learnt by:

$$Z_1 \leftarrow \min_Z \|X - D_1 Z_1\|_F^2 \qquad (9a)$$

$$D_1 \leftarrow \min_D \|X - D_1 Z_1\|_F^2 \qquad (9b)$$

This is the method of optimal directions [41] and both (9a) and (9b) are simple least square problems having closed form solutions.

For the second layer, the substitution is $Z_2 = \varphi(...\varphi(D_N Z))$, which leads to $Z_1 = \varphi(D_2 Z_2)$, or alternately, $\varphi^{-1}(Z_1) = D_2 Z_2$; this too is a single layer dictionary learning. Since the representation is dense, it can be solved using

$$\min_{D_2, Z_2} \|\varphi^{-1}(Z_1) - D_2 Z_2\|_F^2 \qquad (10)$$

This too is can be solved by alternating minimization as in the case of first layer (9). Continuing in this fashion till the penultimate layer, in the final layer one has $Z_{N-1} = \varphi(D_N Z)$ or $\varphi^{-1}(Z_{N-1}) = D_N Z$. In the last level the coefficient Z will be sparse. For learning sparse features, one needs to regularize by applying $l_1$-norm on the features. This is given by:

$$\min_{D_N, Z} \|\varphi^{-1}(Z_{N-1}) - D_N Z\|_F^2 + \lambda \|Z\|_1 \qquad (11)$$

This is solved using alternating minimization.

$$Z \leftarrow \min_{Z} \|\varphi^{-1}(Z_{N-1}) - D_N Z\|_2^2 + \lambda \|Z\|_1 \qquad (12a)$$

$$D_N \leftarrow \min_{D_N} \|\varphi^{-1}(Z_{N-1}) - D_N Z\|_F^2 \qquad (12b)$$

As before, (12b) is a least square problem having a closed form solution. The solution to (12a), although not analytic, can be solved using the Iterative Soft Thresholding Algorithm (ISTA) [42].

*B. Discriminative Robust Deep Dictionary Learning*

Dictionary learning employs the Euclidean ($l_2$-norm) cost function; this is mainly because it has a closed form solution (easy to minimize). It is well known that the Euclidean norm is sensitive to sparse but large outliers. The $l_2$-norm minimization works when the deviations are small – approximately Normally distributed; but fail when there are large outliers. As mentioned before hyperspectral images are corrupted by a mixture of Gaussian and sparse noise. The overall noise has a heavy tailed distribution. For such cases, the Euclidean norm is not ideal.

In statistics there is a large body of literature on robust estimation. The Huber function [43] has been in use for more than half a century in this respect. The Huber function is an approximation of the more recent absolute distance based measures ($l_1$-norm). Recent studies like [44] in robust estimation prefer minimizing the $l_1$-norm instead of the Huber function. The $l_1$-norm does not bloat the distance between the estimate and the outliers and hence is more robust (compared to $l_2$-norm). One can employ the $l_p$-norm (0<p<1) to get more robust estimates, but it would make the problem non-convex. Hence the convex $l_1$-norm is preferred over the $l_p$-norm.

Adding robustness to the dictionary learning problem is the first improvement from [18]. The second (and major) addition is the employment of a discriminative penalty. This allows us to learn a classifier from the last representation layer to the targets. Hence, we have an in-built classifier. This would preclude use of third party classifiers as required in prior deep dictionary learning studies [17, 18].

*1) Training*

In a robust learning approach, during the training phase we intend to solve the following:

$$\min_{D_1,\ldots,D_N,Z} \|X - D_1 \varphi(D_2 \varphi(\ldots\varphi(D_N Z)))\|_1 + \lambda \|Z\|_1 \qquad (13)$$

Note the difference from the DDL formulation; instead of the usual Euclidean norm we have an $l_1$-norm for robust dictionaries. The second one is for learning sparse coefficients.

The aforesaid form (13) does not yet comprise of the discriminative penalty. We incorporate this in the following (final) formulation.

$$\min_{D_1,\ldots,D_N,Z,W} \|X - D_1 \varphi(D_2 \varphi(\ldots\varphi(D_N Z)))\|_1 + \lambda \|Z\|_1 \\ + \mu \|T - WZ\|_F^2 \qquad (14)$$

Here T are the targets, i.e. binary codes for class labels; it has one in the position of the class and zeroes elsewhere. W is the discriminative linear map from the deepest representation layer to the target.

Solving the problem (14) exactly is difficult. First, it is non-convex and second it is computationally demanding. As is typical in deep learning, we follow a greedy approach, i.e. for the first layer, express: $Z_1 = \varphi(D_2 \varphi(\ldots\varphi(D_N Z)))$; so that the shallowest (first) layer of dictionary learning in (14) can be expressed as,

$$X = D_1 Z_1$$

A greedy approximate solution can therefore be formulated as,

$$\min_{D_1, Z_1} \|X - D_1 Z_1\|_1 \qquad (15)$$

Sparsity or the discriminative term does not have any effect on the first layer while learning greedily.

This (15) is the robust (single layer) dictionary learning formulation. In this work, we follow the Split Bregman approach outlined in [45] to solve it. We introduce a proxy variable: $P = X - D_1 Z_1$. The equality between the proxy and the actual variables need not be enforced in every iteration; the constraint needs to be enforced only at convergence. Therefore to relax the equality constraint we introduce the Bregman relaxation variable (B) between the proxy and the actual variable, the Augmented Lagrangian becomes:

$$\min_{P, D_1, Z_1} \|P\|_1 + \mu \|P - X - D_1 Z_1 - B\|_F^2 \qquad (16)$$

This can be segregated into the following subproblems:

$$\min_{P} \|P\|_1 + \mu \|P - X - D_1 Z_1 - B\|_F^2 \qquad (17)$$

$$\min_{D_1} \|P - X - D_1 Z_1 - B\|_F^2 \qquad (18)$$

$$\min_{Z_1} \|P - X - D_1 Z_1 - B\|_F^2 \qquad (19)$$

Sub-problems (18) and (19) are straightforward least squares problems having analytic solution in the form of pseudo-inverse. Sub-problem (17) is an $l_1$-minimization problem having a closed form solution via soft thresholding [46]. The last step is to update the Bregman relaxation variable by gradient descent.

$$B \leftarrow P - X - D_1 Z_1 - B \qquad (20)$$

This concludes the derivation for solving (15). Once the coefficients for the first layer are learnt, one can learn the second layer as a single layer of dictionary learning by substituting $Z_2 = \varphi(D_3 \ldots \varphi(D_N Z))$,

$$Z_1 = \varphi(D_2 Z_2) \equiv \varphi^{-1}(Z_1) = D_2 Z_2 \qquad (21)$$

Computing $\varphi^{-1}$ is trivial since it is an element-wise operation. The second level of dictionary and coefficients are solved by minimizing the Euclidean distance. It should be borne in mind that the effects of outliers are removed in the first layer; therefore there is no need to employ the computationally expensive $l_1$-norm minimization in subsequent layers.

$$\min_{D_2, Z_2} \left\| \varphi^{-1}(Z_1) - D_2 Z_2 \right\|_F^2 \quad (22)$$

This is easily solved using alternating minimization. In the $k^{th}$ iteration –

$$Z_2(k) \leftarrow \min_Z \left\| \varphi^{-1}(Z_1) - D_2(k-1)Z \right\|_F^2 \quad (23a)$$

$$D_2(k) \leftarrow \min_D \left\| \varphi^{-1}(Z_1) - DZ(k-1) \right\|_F^2 \quad (23b)$$

The same greedy process can be continued to deeper layers till the penultimate layer. In the final layer, we will have

$$Z_{N-1} = \varphi(D_N Z) \equiv \varphi^{-1}(Z_{N-1}) = D_N Z \quad (24)$$

Noting that the coefficients in the final layer should be sparse; one also need to solve for the discriminative linear map. The optimization problem is formulated as:

$$\min_{D_N, Z, W} \left\| \varphi^{-1}(Z_{N-1}) - D_N Z \right\|_F^2 + \lambda \|Z\|_1 + \mu \|T - WZ\|_F^2 \quad (25)$$

This too can be solved using alternating minimization.

$$Z(k) \leftarrow \min_Z \left\| \varphi^{-1}(Z_{N-1}) - D_N(k-1)Z \right\|_F^2 + \lambda \|Z\|_1 + \mu \|T - WZ\|_F^2 \quad (26a)$$

$$D_N(k) \leftarrow \min_D \left\| \varphi^{-1}(Z_1) - DZ(k-1) \right\|_F^2 \quad (26b)$$

The dictionary update remains the same as before. Sub-problem (26a) can be expressed as follows:

$$\min_Z \left\| \begin{pmatrix} \varphi^{-1}(Z_{N-1}) \\ \sqrt{\mu} T \end{pmatrix} - \begin{pmatrix} D_N(k-1) \\ \sqrt{\mu} W \end{pmatrix} Z \right\|_F^2 + \lambda \|Z\|_1 \quad (27)$$

This is an $l_1$-regularized least squares problem. It can be solved using Iterative Soft Thresholding Algorithm (ISTA) [42].

This concludes the derivation. It is a greedy approach; therefore there is no feedback between layers. We will show, even without the fine-tuning (feedback) our method yields better results than other fine-tuned deep learning tools.

The value of $\mu$ is kept to be unity in this work. This is because, we give equal importance to both representation learning and classification. The parameter $\lambda$ needs to be specified by the user.

*2) Robust Deep Dictionary Learning – Testing*

During testing, the first task is to generate the representation for a new test sample – $x$. We need to solve:

$$\min_z \left\| x - D_1 \varphi \left( D_2 \varphi (\ldots \varphi (D_N z)) \right) \right\|_1 + \lambda \|z\|_1 \quad (28)$$

Owing to the non-linearity, solving (28) in a straight-forward fashion is not easy. Therefore we resort to a greedy technique. We learn the representation in layers; for the first layer, this is $z_1 = \varphi \left( D_2 \varphi (\ldots \varphi (D_N z)) \right)$. Greedy substitution leads to solution of $z_1$ via

$$\min_{z_1} \|x - D_1 z_1\|_1 \quad (29)$$

There are many techniques to solve (29). Here we use the non-parametric iterative re-weighted least squares technique.

In the second level, the substitution is $\varphi^{-1}(z_1) = D_2 z_2$ where $z_2 = \varphi(\ldots \varphi(D_N z))$. The representation at the second layer is solved using simple least squares (because the outliers are removed in the first layer):

$$\min_{z_2} \left\| \varphi^{-1}(z_1) - D_2 z_2 \right\|_2^2 \quad (30)$$

The substitution continues till the penultimate level. In the final level, the problem we need to solve is,

$$\min_z \left\| \varphi^{-1}(z_{N-1}) - D_N z \right\|_F^2 + \lambda \|z\|_1 \quad (31)$$

This is the standard $l_1$-minimization problem. We solve it using the ISTA [42]. The representation from the final layer ($z$) is used for classification.

Once the representation is learnt, we need to classify it. This is done by multiplying the representation $z$ by the learnt classifier map. This gives us the target: $t = Wz$. Ideally it should contain a 1 in one of the positions and 0's elsewhere; however such is not the case in practice. To get the class of the test sample, we seek the position of the highest coefficient in $t$ – this gives us the class label.

## IV. EXPERIMENTAL EVALUATION

We evaluate our proposed technique on the problem of hyperspectral image classification; the datasets are Indian Pines which has 200 spectral reflectance bands after removing bands covering the region of water absorption and 145*145 pixels of sixteen categories, and the Pavia University scene which has 103 bands of 340*610 pixels of nine categories.

TABLE I
TRAINING AND TEST SAMPLES FOR INDIAN PINES

| Class | Training Samples | Test Samples | Total Samples |
|---|---|---|---|
| 1 | 15 | 31 | 46 |
| 2 | 142 | 1286 | 1428 |
| 3 | 83 | 747 | 830 |
| 4 | 23 | 214 | 237 |
| 5 | 48 | 435 | 483 |
| 6 | 73 | 657 | 730 |
| 7 | 20 | 8 | 28 |
| 8 | 47 | 431 | 478 |
| 9 | 15 | 5 | 20 |
| 10 | 97 | 875 | 972 |
| 11 | 160 | 2295 | 2455 |
| 12 | 59 | 534 | 593 |
| 13 | 20 | 185 | 205 |
| 14 | 126 | 1139 | 1265 |
| 15 | 38 | 348 | 386 |
| 16 | 50 | 43 | 93 |

TABLE II
TRAINING AND TEST SAMPLES FOR PAVIA UNIVERSITY

| Class | Training Samples | Test Samples | Total Samples |
|---|---|---|---|
| 1 | 132 | 6499 | 6631 |
| 2 | 372 | 18277 | 18649 |
| 3 | 41 | 2058 | 2099 |
| 4 | 61 | 3003 | 3064 |
| 5 | 26 | 1319 | 1345 |
| 6 | 100 | 4929 | 5029 |
| 7 | 26 | 1304 | 1330 |
| 8 | 73 | 3609 | 3682 |
| 9 | 18 | 929 | 947 |

Prior studies on deep learning based classification assumed an overtly optimistic scenario [2 – 4] – they assumed 80% (60% training + 20% validation) labelled data is available; and only 20% need to be predicted. This is an unrealistically favorable protocol. In this work we follow the more standard evaluation protocol on these datasets. For the first dataset (Indian Pines), we randomly select 10% of the labelled data as training set and rest as testing set; for the second dataset (Pavia University) 2% of the labelled data is used for training and the rest for testing.

For the prior DDL based formulation [18] a neural network is used for classification. Our proposed formulation (DRDDL) and other deep learning architectures [2-4] have in-built classifiers. Both DDL and our proposed DRDDL, have a 3 layer deep architecture of 150-100-30 for Indian Pines and 3 layer deep architecture of 80-40-20 for Pavia University. The architecture differs owing to the volume of the training data. Since Indian Pines has more training samples it can afford a larger architecture (more basis). The same architecture overfits in case of Pavia; so we defined a smaller one. The value of the sparsity penalty $\lambda$ has been fixed to 0.2 for all the problems.

In the first set of experiments the input consists of raw data of all the spectral channels pixel-wise (spatial features). We compare with [2-4] in the given protocol and report the results in Table III. The performance is measured in terms of the three standard measures – overall accuracy (OA), average accuracy (AA) and kappa.

TABLE III
CLASSIFICATION WITH RAW PIXEL VALUES

| Dataset | Metric | DRDDL | DDL [18] | SDAE [2] | DBN [3] | CNN [4] |
|---|---|---|---|---|---|---|
| Pavia | AA | **94.79** | 91.34 | 81.03 | 74.88 | 69.91 |
|  | OA | **96.98** | 92.51 | 87.89 | 78.06 | 74.57 |
|  | Kappa | **0.96** | 0.93 | 0.88 | 0.80 | 0.76 |
| Indian Pines | AA | **73.67** | 70.76 | 67.78 | 65.69 | 59.29 |
|  | OA | **82.11** | 77.84 | 70.23 | 67.38 | 63.50 |
|  | Kappa | **0.85** | 0.78 | 0.71 | 0.66 | 0.65 |

TABLE IV
COMPARISON WITH BEST IN CLASS TECHNIQUES

| Dataset | Metric | DRDDL | DDL [18] | SDAE [2] | DBN [3] | CNN [4] | [33] |
|---|---|---|---|---|---|---|---|
| Pavia | AA | **98.11** | 92.67 | 85.02 | 78.50 | 87.12 | 90.39 |
|  | OA | **98.29** | 94.56 | 88.26 | 86.09 | 95.34 | 97.33 |
|  | Kappa | **0.98** | 0.93 | 0.90 | 0.84 | 0.94 | 0.97 |
| Indian Pines | AA | **87.45** | 86.98 | 78.33 | 73.33 | 83.19 | 86.77 |
|  | OA | **93.08** | 90.03 | 86.10 | 81.79 | 90.21 | 88.33 |
|  | Kappa | **0.86** | 0.83 | 0.73 | 0.67 | 0.78 | 0.80 |

The results show that both DDL and the proposed RDDL yields considerably superior results compared to the existing deep learning techniques. However, one cannot compare the results shown here with [2-4]. This owes to two reasons – first there is no pre-processing here, and second, the training to testing ratio is more realistic than used in the aforesaid papers. Robust deep dictionary learning supersedes deep dictionary learning for reasons discussed before; the inbuilt robustness combats the mixed noise inherently present in the hyper-spectral data better than ordinary dictionary learning.

In the final set of results (Table IV) we compare the best techniques reported in [2-4] along with the proposed pre-processing, feature extraction and classification. In [2] spatial features in terms of patches are concatenated with spectral features obtained by PCA as inputs for stacked autoencoders. The outputs of the autoencoders are used for classification via logistic regression. In [3] the authors use the same features; instead of inputs to SAE they input to DBN. The rest remains the same. In [4] CNNs are trained enforcing sparsity.

We also compare with another alternative CNN based formulation in this area [33] that uses a combination of CNN and balanced local discriminant embedding (BLDE) for feature extraction followed by fusion at the feature and classifier level to yield the final classification result.

Owing to its simplicity and effectiveness we follow the feature extraction scheme of [2, 3]; the other deep learning techniques [4, 33] use CNN and hence the feature extraction is amenable to our technique. Following [2, 3], we extract patches for spatial information and PCA for spectral information. These are concatenated to form the final feature vector. This is in turn fed into deep dictionary learning for feature extraction. For classification we employ the kernel sparse representation based classifier [47].

The results (Table IV) show that both DDL and Robust DDL yield superior results than recent deep learning based classification techniques. The studies [2-4] are significantly worse than DDL based methods. The most recent work [50] is better compared to the rest [2-4] but is worse than DDL and Robust DDL.

The reason [33] does better than existing deep learning techniques is because it uses pre-trained CNN models. Usual deep learning models are sensitive to the number of training samples. The number of training samples we have used here are drastically smaller (by an order of magnitude) than the ones used in deep learning papers. Hence the prior studies [2-4] suffer. But owing to unsupervised pre-training, [33] is able to combat the curse of limited training samples to a certain extent.

For visual evaluation, we show the classification results from different techniques. In Fig. 9 we show results from the Pavia dataset on raw pixel values. This corresponds to Table III. Fig. 10. corresponds to Table IV; here we show comparison with the best in class techniques on the Indian Pines dataset. For both Fig. 9 and 10, the images corroborate the numerical results.

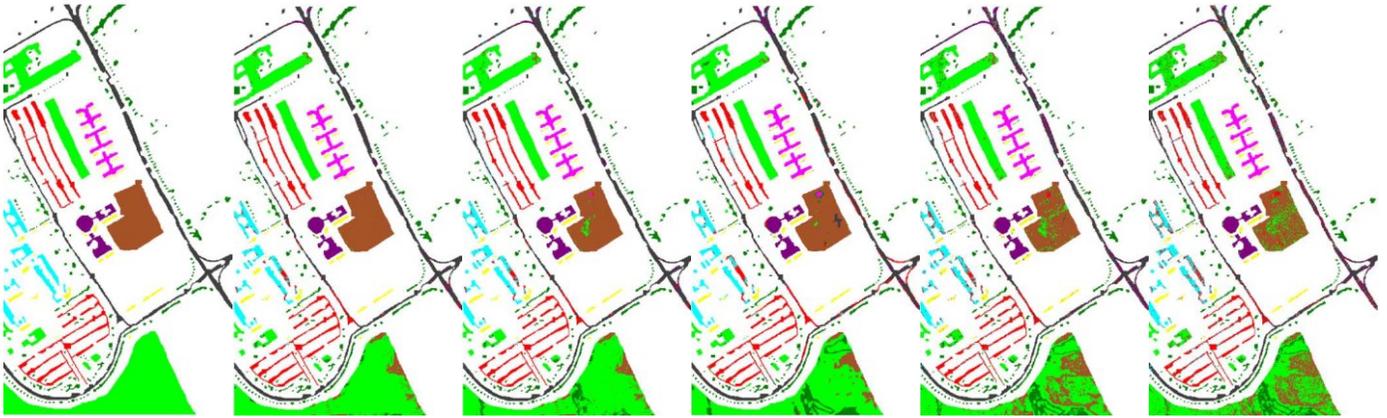

Figure 9. Pavia. Raw Pixel Values. Left to Right – Groundtruth, DRDDL, DDL, SDAE [2], DBN [3], CNN [4]

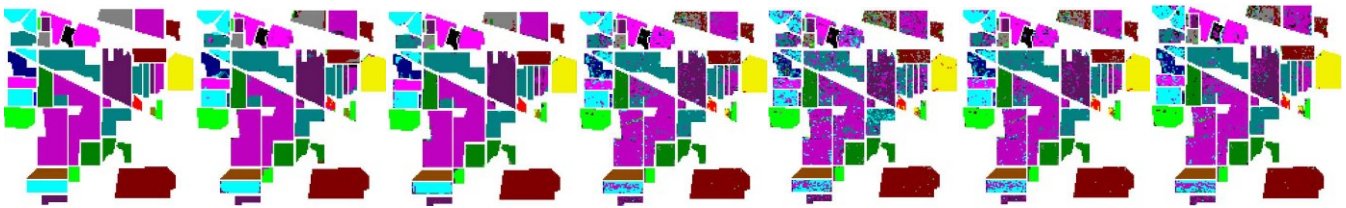

Figure 10. Indian Pines. Best in Class. Left to Right – Groundtruth, DRDDL, DDL, SDAE [2], DBN [3], CNN [4] and CNN [33]

## V. CONCLUSION

In this work we address the problem of hyperspectral image classification. In recent years there are a quiet a few comprehensive studies on these topics [2-4, 33]; these are straightforward applications of deep learning tools like stacked autoencoder [2], deep belief network [3] and convolutional neural network [4] on hyperspectral datasets. A fourth framework for deep learning has been recently proposed by the authors [17]. This is deep dictionary learning. Here multiple levels of dictionaries are learnt to represent the data. Our work is based on the deep dictionary learning framework.

The first work applying deep dictionary learning for hyperspectral image classification is [18]. This paper improves in two ways. First, instead of employing the usual $l_2$-norm cost function, we incorporate the more robust $l_1$-norm. This is especially suitable for hyperspectral imaging problems, since they are known to be corrupted by a mixture of Gaussian and sparse noise. The second improvement is the incorporation of a discriminative linear map. This allows us to have an inbuilt classifier into the deep dictionary learning framework; it does not require a separate classifier like [17, 18].

For hyperspectral imaging we have carried out experiments with all possible variants deep learning [2-4, 33]. In all cases, we find that our method to be superior to existing ones.

Previously most studies in dictionary learning based solutions to inverse problems applied redundant dictionaries while problems in analysis / classification employed under-complete dictionaries. In recently times is a concerted effort to build orthogonal dictionaries [48, 49] especially for inverse problems. It remains to be seen, if such techniques can be employed for analysis as well.